\documentclass{emulateapj}

\usepackage{graphicx}
\usepackage{bm}
\usepackage{psfrag}
\usepackage{mathrsfs}
\usepackage{placeins}
\usepackage{amssymb,amsmath}
\newcommand{\be}{\begin{eqnarray} }
\newcommand{\ee}{ \end{eqnarray} }

\shorttitle{Radio emission of PSR~J0901$-$4624}
\begin{document}
\title{Two radio-emission mechanisms in PSR~J0901$-$4624}

\shortauthors{C.~A.~Raithel et al.}
\author{C.~A.~Raithel\altaffilmark{1,2}, R.~M.~Shannon\altaffilmark{1,3}, S.~Johnston\altaffilmark{1}, \& M.~Kerr\altaffilmark{1}}
\affil{$^1$ CSIRO Astronomy and Space Science, Australia Telescope National Facility, Box 76 Epping, NSW 1710, Australia,}
\affil{$^2$ Department of Physics and Astronomy, Carleton College, Northfield, MN  55057 USA}
\altaffiltext{3}{E-mail:  ryan.shannon@csiro.au}




\begin{abstract}
We have detected  sporadic, bright, short-duration radio pulses from 
PSR~J0901$-$4624. 
These pulses are emitted simultaneously with persistent, periodic emission that dominates the flux density when averaging over many periods of the pulsar.
The bright pulses have energies that are consistent with  a power-law distribution.
The integrated profile of PSR~J0901$-$4624 is highly polarized and 
shows four distinct components. The bright pulses appear to originate
near the magnetic pole of the pulsar and have polarization properties unlike
that of the underlying emission at the same pulse phase.
We conclude that the bright pulses represent a secondary giant-micropulse emission  process, possibly from a different region in the  pulsar magnetosphere.  
\end{abstract}


\section{Introduction}

Since the discovery of pulsars, the phenomenology
associated  with their  radio emission has been a fertile source for
studying  coherent-emission processes in magnetized plasma.
It is clear that although the time-integrated
profiles of radio pulsars are constant or nearly constant,  individual pulses that go
into creating the integrated profile are highly variable both in 
total intensity and in their polarization.

Historically, there has been an understanding that the fluence   of
individual pulses follows either a  log-normal or  Gaussian distribution
\citep{cjd01,cjd03a,bjb+12,sod+14}
with relatively few pulses significantly stronger
than the mean.  We will refer to this distribution as the normal-mode of emission.  
The main exception to this has been the giant pulses
observe from the Crab pulsar \citep{sr68}, which are extremely bright ($ \gtrsim 1$~MJy) and
 narrow ($\ll 10$~ns) bursts  emitted at certain pulse phases \cite[][]{hkw+03}.

More recently,  a sub-set of pulsars have been identified that show
highly erratic behaviour in their single pulses. 
The giant-pulse mode was
seen in other young \citep{jr03} and millisecond pulsars
\citep{cst+96,rj01,kbm+06}.
Several young, energetic pulsars show single-pulse variability  similar to the giant-pulse mode,
 except that the bursts have $\mu$s rather than ns
widths \citep{jvkb01,jr02}, which has been referred to as the giant-micropulse emission.

Emission from less energetic pulsars and millisecond pulsars also shows phase dependent variability, including power-law statistics on the leading or trailing edges.

One of the best-studied slow pulsars, PSR~B1133$+$16  shows, at high frequencies ($\nu > 4$~GHz) individual pulses with power-law statistics, of $\sim$ 2 ms width \cite[][]{2003A&A...407..655K,2014MNRAS.440..457K}.
The bright pulses  originate predominantly from the trailing edge of the first component of the double-cone profile.
At low frequencies, the bright pulses  originate on both components  \cite[][]{2003A&A...407..655K}. 
  The brightest pulses from the millisecond pulsar PSR~J1713$+$0747 originate from the trailing edge of the main component \cite[][]{2012ApJ...761...64S}.  However these pulses show a lognormal energy distribution. 
  In contrast, the brightest pulses in PSR J0437$-$4715 originate from the center of the profile.  
These stronger pulses have fractionally higher polarization than weaker pulses \cite[][]{2014MNRAS.441.3148O}.

Rotating Radio Transients \cite[RRATs,][]{mll+06}
emit single pulses extremely sporadically, with pulses emitted every few minutes or longer.
While some of the pulsars in this group also have underlying normal-mode 
emission, some do not \cite[][]{r10}.
Further complicating the picture,  several pulsars appear to have a combination of normal-mode, and RRAT-like behaviour \citep{wsrw06,wje11}.


Here we report the discovery of a new example of sporadic, 
bright emission from PSR~J0901$-$4624. 
Bright pulses from 
J0901$-$4624 show a power-law distribution, which are 
weaker and broader than giant pulses but  occur more frequently than  giant micropulses identified from previous pulsars.
In addition,
we compare the bright pulses of J0901$-$4624 to other modes of pulsar emission.

\section{Observations}
\label{sec:obs}

During a monitoring programme of  $\sim 180$ young pulsars   at the $64$-meter Parkes radio telescope  in support of the Fermi Space Telescope mission \cite[][]{wjm+10}, using a real-time transient search  algorithm \cite[][]{bbb+12b}, we identified occasional (once per $20$\,s) bright (S/N $> 20$), narrow ($< 500$~$\mu$s) pulses being emitted from PSR~J0901$-$4624. 
With a rotation period of $442$~ms, a dispersion measure
of 199~pc\,cm$^{-3}$, and an inferred spin-down energy of $4.0 \times 10^{34}$~erg\,s$^{-1}$, PSR~J0901$-$4624 has median properties for the  sample selected for high spin-down energy.   

To examine the bright pulses in more detail,  we observed PSR~J0901$-$4624 again with the Parkes telescope on six 
separate occasions,  of $30-60$\,min in duration, as listed in Table \ref{tab:obs}.  All observations were made using the $21$-cm multibeam system at a central frequency  close to  $1400$~MHz. 
The system-equivalent flux density of the system is $\sim35$\,Jy on cold regions of the sky.

Data were recorded using three backend systems operating in parallel.
The first backend (PDFB4)
produced a period-averaged digital-filterbank output of $256$\,MHz bandwidth, centered at $1369$\,MHz, comprising $1024$ frequency channels for
each of $1024$ phase bins across the pulse period for both the two auto correlations
and the real and imaginary parts of the cross correlations of the feed probes.
Data were folded at the topocentric pulse period and accumulated for
$30$\,s and recorded to disk.
The second system (PDFB3) produced fast-sampled digital-filterbank output with $256$~MHz bandwidth centered at $1369$~MHz, 
with $512$~frequency channels per polarization sampled every 256~$\mu$s.
Finally, on May~25 we used a baseband recording system (CASPSR), with $400$~MHz bandwidth centered at $1382$~MHz in order 
to resolve short-duration signals  and mitigate intra-channel dispersion smearing in observations  with the filterbank-spectrometer systems.
The {\sc dspsr} software package \cite[][]{vb11} was used to subdivide the data into single rotations of the pulsar for the PDFB3 and CASPSR datasets. In the case of CASPSR,  {\sc dspsr}  was also used to coherently disperse the data. 

A calibration signal, injected into the feed
at an angle of 45$^\circ$ with respect to the probes, was recorded before each observation.
This signal was used to determine the relative gain  and phase of the two
polarization probes.

Data editing and calibration were conducted using the {\sc psrchive} package \citep{hvm04}.
In brief, after removal of radio-frequency interference in both the frequency and
time domains, the data were gain and polarization calibrated and
summed in frequency and time to produce a pulse profile for each
pulsar observed. 
Data were flux calibrated using observations of the radio galaxy Hydra~A made as part of the Parkes Pulsar Timing Array project \cite[][]{mhb+13}.
Details of the observations can be found in Table \ref{tab:obs}.

\section{Integrated profile}
\begin{figure}

\includegraphics[height=0.4\textheight,angle=0]{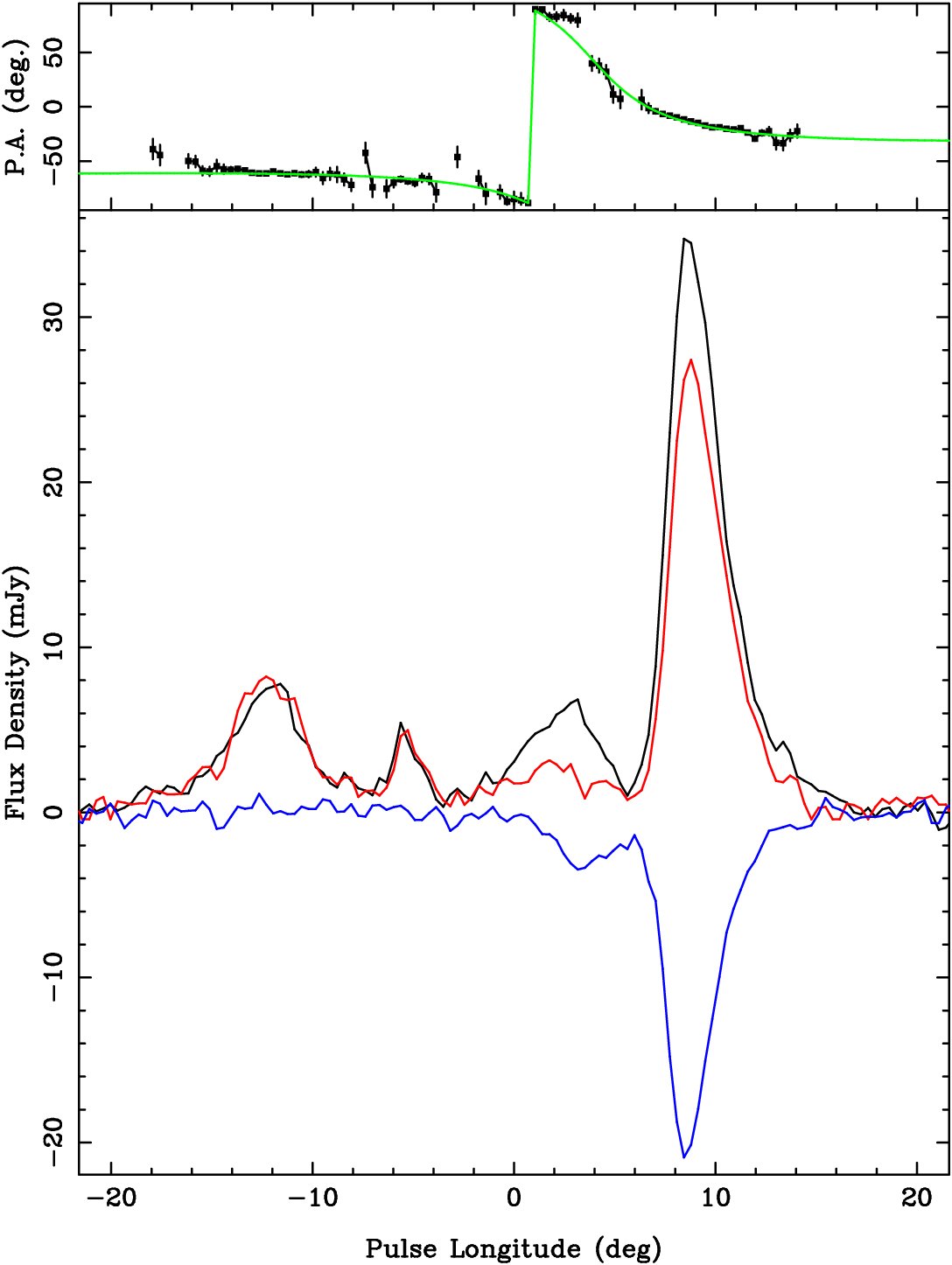} \\
\includegraphics[height=0.4\textheight,angle=0]{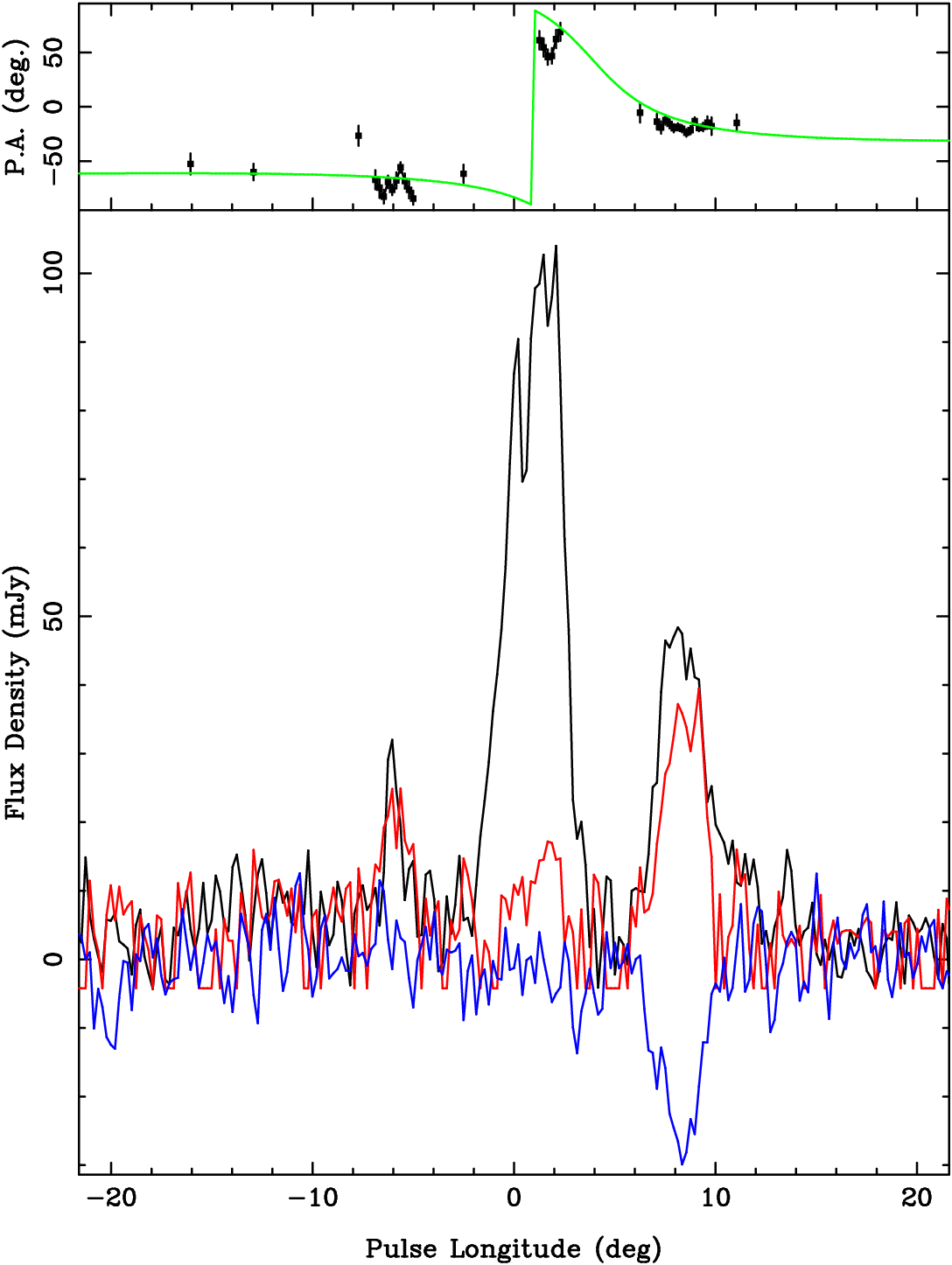}
\caption{Upper panel:  Integrated profile of PSR~J0901$-$4624 in full Stokes at 1369~MHz
The lower plot shows the total intensity (black line), linear polarization
(red line) and circular polarization (blue line) as a function
of pulse longitude. The top panel shows the position angle of the linearly 
polarized radiation and runs between $-90$\degr\ and +90\degr. 
The green line is the best-ftting model of the beam geometry to the position-angle data.
Phase zero
is chosen to be the geometric center of the profile.  The dominant features in the profile are  four components centered at longitudes of $-12\degr$, $-6\degr$, 4$\degr$, and $9\degr$ relative to the geometric center.  Lower panel:  Profile formed from only the bright (S/N $>$ 5) pulses.  Bright pulses originate primarily from the central component and have a width of $1$-$2$ bins of phase resolution.    The RVM model is the same as in the upper panel.}
\label{fig:profile}
\end{figure}

In the top panel of Fig~\ref{fig:profile}, we show the full-Stokes profile and linear polarization position angle 
 for PSR~J0901$-$4624, formed from co-adding all of our observations.
 The profile consists of four~distinct components
with a total width of $27\degr$. 
The geometric center of the profile 
occurs between the two inner components, which indicates the pulsar has
double-conal structure.
The pulsar is highly linearly polarized,  which is typical for young, high-$\dot{E}$ pulsars.  
PSR~J0901$-$4624 also shows a high degree of circular polarization in
the trailing component \citep{jw06,wj08}; it is common for the trailing component to have the dominant amplitude
and significant circular polarization \cite[][]{jw06}.

The position-angle swing is well described by the rotating-vector model with the best-fitting model giving an inflexion point located 3.9\degr\ after the geometric center of the profile.
This can be used to infer an emission height of $\sim$700~km, or about 3\%
of the light-cylinder radius.
Using the method of Rookyard et al. (2014) we determine a best-fitting value
for the inclination of the rotation axis to the magnetic axis of
$\alpha \sim$50\degr\ and an impact parameter of $\beta \sim$2\degr.

Observations of the pulsar have also been made at $3.1$~GHz and $0.7$~GHz. There
is little evolution of the profile over this frequency range.

\section{Individual Pulses}

In total, over the six observations we observed $\sim$34,000 rotations of the pulsar.  
Approximately $\sim$400   pulses with a S/N $>$~5 were identified.  All  of these bright pulses had narrow width ( $< 1$~ms).
The daily variation in the rate of bright pulses, as shown in Table \ref{tab:obs} is caused by scintillation and the distribution of pulse energies (see section \ref{sec:fluxdistro}).  After accounting for these effects, consistent numbers of bright pulses were detected on each day.  

\begin{deluxetable}{crrrr}
\tabletypesize{\scriptsize}
\tablecaption{Observation Details}
\tablehead{ \colhead{Date}  & \colhead{$N$} & \colhead{$N_5$}  &\colhead{  $\bar{S}_\nu$}  \\
  \colhead{(MJD)} &  & &  \colhead{(mJy)}} 
 \startdata
May 17 & 8148 & 32 & 0.66   \\  
May 25 &  8156 & 33& 0.69\\
July 25 & 7034  & 72& 0.95\\
July 27 & 2197 & 17& 0.85\\
Aug 19 & 4085 & 13& 0.63\\
Sep 28 & 4085 & 5 &0.51 \enddata
\label{tab:obs}
\tablecomments{For each epoch, we list the $N$, number of rotations of the pulsar observed; $N_5$, the number of $5-\sigma$ pulses detected, and $\bar{S}_\nu$, the mean observed flux density.}
\end{deluxetable}

\begin{figure}
\begin{tabular}{cc}
\multicolumn{2}{c}{\includegraphics[scale=0.4,angle=0]{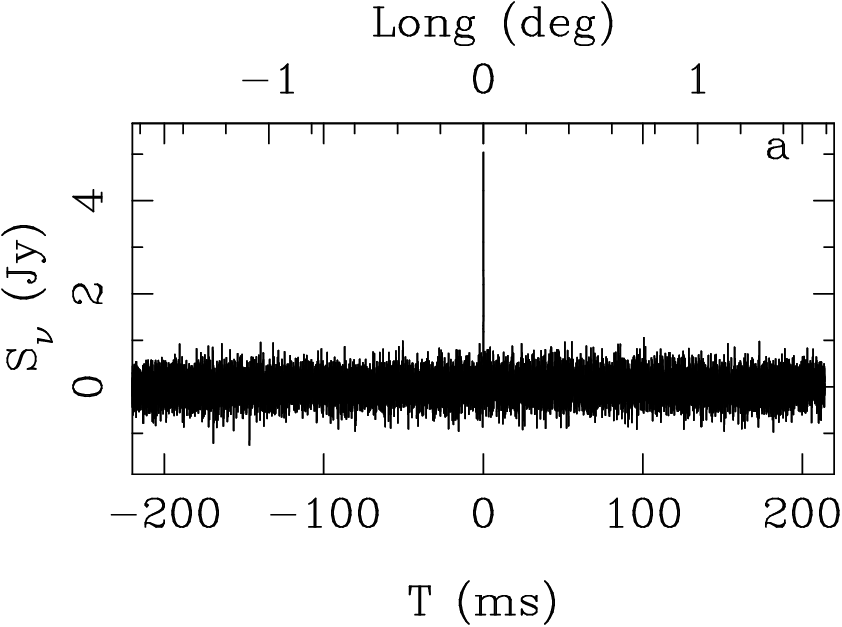} }\\
\includegraphics[scale=0.4,angle=0]{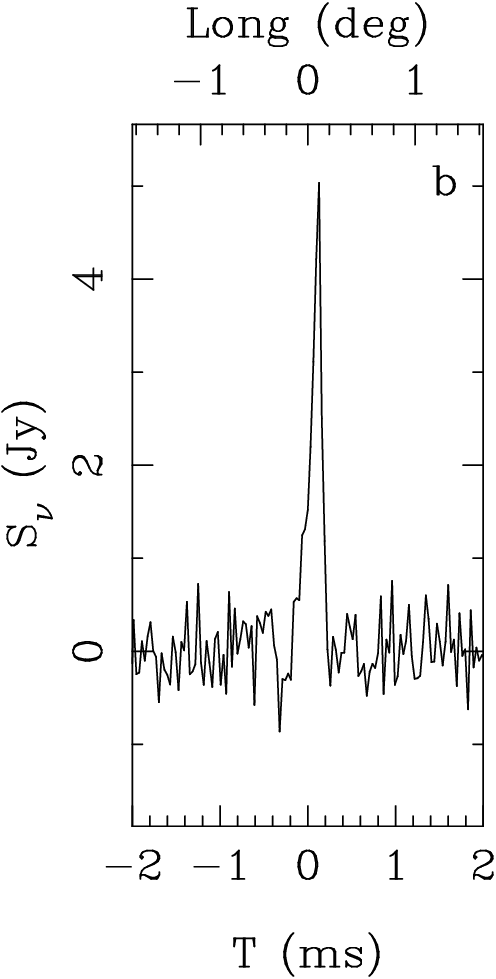} & \hspace{0.1in}  \includegraphics[scale=0.4,angle=0]{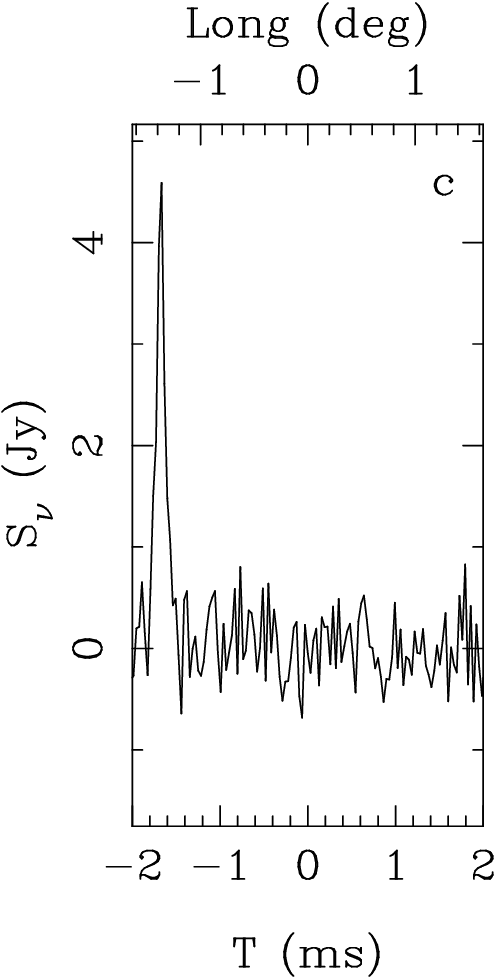}\\
\end{tabular}
\caption{Two bright  pulses taken from our coherently dedispersed observations with CASPSR.   The data have been binned to provide 32~$\mu$s resolution.  Panel {\em a}  Full rotation of PSR J0901$-$4624.  Panels {\em b} zoom-in of pulse displayed in Panel {\em a}.  Panel {\em c}:   Second bright pulse.    }
\label{fig:caspsr}
\end{figure}

Based on observations made with the CASPSR coherent dedispersion system, the full-width half maximum of the bright pulses varies between 130~$\mu$s and 400~$\mu$s, which is less than $0.1$\% of the pulse period.
Two of the brightest pulses from these observations are displayed in Figure \ref{fig:caspsr}.
The location of the bright pulses varies in location by  approximately  4\degr\ in pulse phase (1\% of the period).
Occasionally bright pulses are seen from the two components that flank the central component.
The time between bright pulses is consistent with a Poisson random process.

In the lower panel of Figure \ref{fig:profile}, we show the average profile formed by combining only the brightest pulses. 
These pulses greatly enhance the emission at the center of the integrated profile.
We note that the rarity of bright pulses results in very little overall effect on the integrated profile.
In addition, we see the bright pulses are emitted simultaneously with the underlying,
fainter  emission. 
The normal emission observed during the bright pulses  ($S_{\rm peak, bright} \sim 50$~mJy) is modestly brighter 
during all of the pulses ($S_{\rm peak, all} \sim 35$~mJy) suggesting that there may be a connection between the two modes of emission.

\subsection{Polarization Properties}

The integrated profile of PSR~J0901$-$4624 is highly linearly polarized (top panel of Figure~1). This implies that one orthogonal mode dominates the emission \cite[][]{2005MNRAS.364.1397J} and that the single pulses should also be highly polarized. 
For the normal emission this is indeed the case.  We occasionally detect bright pulses from component 2 and 4 which are highly polarized (see lower panel of Figure~1).

In contrast, the bright emission has very different polarization as can be seen in Figure 1. 
In the centre of the profile, where the bright emission originates, almost no polarization is seen and what little polarization there is has a different polarization-angle swing to that of the normal emission. 
This is unusual behaviour and hints at a different origin for the normal and bright emissions.

\section{Flux distribution} \label{sec:fluxdistro}

\begin{figure}
\includegraphics[width=0.45\textwidth]{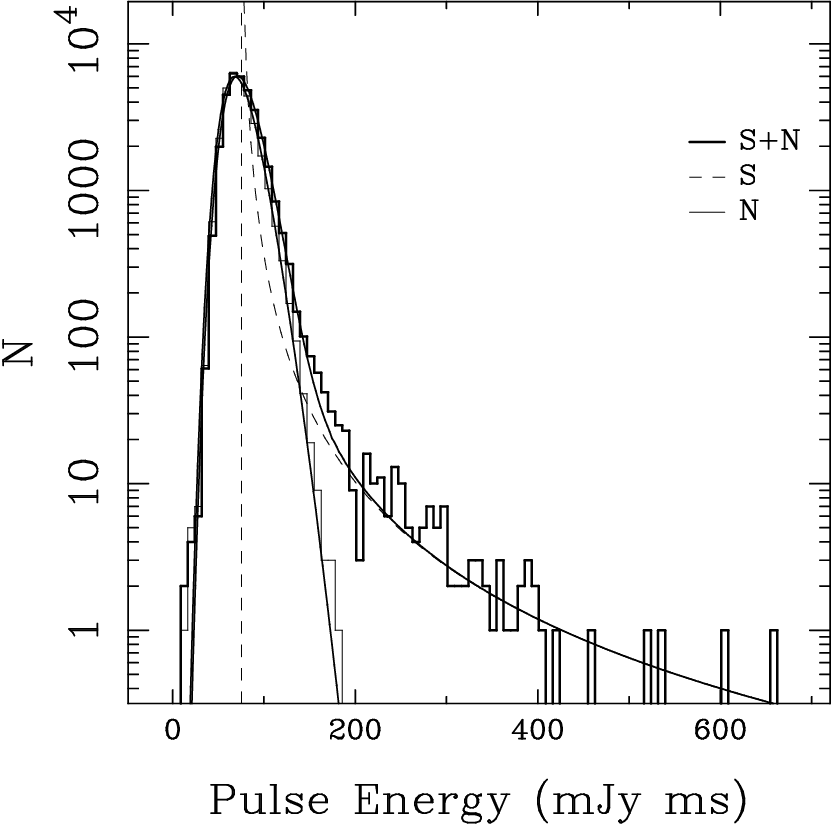}
\centering
\caption{Pulse-energy distribution for PSR~J0901$-$4624.  The thin solid line labelled {\em N} shows a model for the noise.  The thin-dashed line labelled $S$  shows the best-fitting power-law model for the pulse energy distribution.  The thick solid line (labelled $S+N$) shows the convolved pulse-energy distribution.  }
\label{fig:fluxdistro}
\end{figure}

We analyzed the flux distribution using the single-pulse PDFB4 data.  
We identified bright pulses by first identifying the brightest bin in a  window of  $14^\circ$ width in pulse phase    centered on the third component of the average profile.
We then calculated three signal-to-noise (S/N) ratios using an estimate of the off pulse noise:  the first from only the brightest bin, and the second and third from the sum of the brightest and its two adjacent bins.
We only searched for narrow pulses after a first inspection of the data revealed that the bright pulses were either unresolved ($< 1$ bin of pulse phase for the lower time resolution PDFB4 data) or marginally resolved. 
Of these three, we report the integrated pulse energy with the largest S/N.
To assess the expected distribution of noise, we repeated the  same algorithm using a control sample of the same width in a region of pulse phase in which no pulsar emission occurs. 
In Figure \ref{fig:fluxdistro}, we show that the histograms of pulse energy in both the bright pulse window and the off-pulse window.


Because of the relatively low signal-to-noise ratio, it is necessary to deconvolve the noise distribution when modelling the pulse-energy distribution.
We used techniques described in \cite{sod+14} to deconvolve the noise and model the pulse-energy distribution.   The measured pulse energy $\rho_E$ is the convolution of the noise $\rho_N$ and the intrinsic energy $\rho_I$:
\be
\label{eqn:convolution}
\rho_E(E)  = \int dE^\prime \rho_N(E^\prime) \rho_I(E-E^\prime).
\ee

The  noise distribution $\rho_N$ is  non-trivial to calculate because the algorithm used to identify significant pulses results in a non-Gaussian statistics, and refractive scintillation modulates the S/N at each epoch.
For a single observing epoch, in which we assume that the scintillation does not significantly affect the pulse energy,
by the central limit theorem, the off-pulse samples are expected to be Gaussian distributed.
The probability of getting at least one measurement greater than some value $E$ in $N$ trial samples is the converse of the probability of having all values less than $E$
\begin{equation}
\label{eqn:cdf}
P(E > E^\prime) = \frac{1}{2} \left[ 1 + {\rm erf}\left(\frac{E}{\sqrt{2}\sigma} \right) \right]^N.
\end{equation}

The probability density is  the derivative of Equation (\ref{eqn:cdf}):
\be
\rho_N(E) = \frac{N}{2} \left[  1 + {\rm erf}\left(\frac{E}{\sqrt{2}\sigma} \right) \right]^{N-1} \exp\left[-\frac{x^2}{2 \sigma^2} \right]. 
\ee

  Because we have averaged together multiple days, after correcting for pulse flux variations caused by scintillation, the noise distribution deviates from theoretical expectation.  We therefore fitted an analytic function to the off-pulse noise distribution, shown as the thin, solid line in Figure \ref{fig:fluxdistro}, with $N$ and $\sigma$ allowed to be free parameters, and we then deconvolved the pulse-energy distribution using this function. 

For the pulse energy distribution we considered both  log-normal distribution (see Equation 7 of \cite{sod+14}) and  power-law distribution
\be
\rho_I(E)&&=  A E^{-\beta} \exp(-E_c/E) ~~~(E< E_c) \nonumber\\
  &&=  A E^{-\beta}  ~~~~~~~~~~~~~~~~~~~(E  >E_c), 
\ee
where  $\beta$ is the spectral index, $E_c$ is a low-energy exponential cutoff, and $A$ is a normalising factor.

We find that models with power-law distributions produced better agreement than models 
with log-normal distributions.
In particular the best-fitting log-normal models do not reproduce the observed high-energy tail.
For the best-fitting power-law model,  the energy cut-off is $E_c  \approx 0.1 $~mJy~ms and  the spectral index is $\alpha = 2.28(4)$.  The spectral index is comparable to those measured in other power-law emitting pulsars \cite[][]{kbm+06}.  

\section{Discussion}

One of the major open questions in pulsar emission  is how timescales are set for pulse emission. 
 Giant pulses have timescale of ns, whereas the intermittent pulsars \cite[][]{klo+06} can remain in quiescent states for years between active states. 
In between these two extremes are
microstructure, single pulse drifting, nulling, mode-changing and  RRAT emission.
There is clearly a continuum of temporal regimes, but  it is important 
to assess how these variations relate to other pulsar properties  such ages,  viewing and beaming geometries, and magnetic field strengths.   PSR~J0901$-$4624 shows a secondary emission mode in which bright narrow pulses are frequently emitted from the central component.

Another important differentiator between emission processes is the distribution of the energies of individual pulses \cite[see ][ and references within]{cjd01}.
Normal-mode  emission (from slow or millisecond pulsars) is observed to have either log-normal or gaussian energy distributions \cite[][]{bjb+12,sod+14}.
The situation with RRAT-mode emission is much
less clear. 
The best-studied object with RRAT emission, PSR~J1819$-$1458 shows sporadic pulses
that are very similar  to  normal-mode emission \cite[][]{khv+09}.
However, other RRATs appear to have power-law emission \cite[][]{r10}.
In contrast, giant pulses, giant micropulses of Vela, the trailing
bright pulses in PSR~B1706$-$44, and the emission of PSR~J0901$-$4624 follow power-law distributions.

The width of the bright pulses in PSR~J0901$-$4624 seem to
lie between the giant pulses (with widths less than 10$^{-5}$P) and
those of the RRATs (with widths greater than 10$^{-3}$P).  The jitter
in the arrival times is also greater than in  total pulse phase giant-pulse emission, less than RRAT-mode emission, and similar to giant micropulse emission.  The widths of the pulses can be connected to the  physical scales of the emission region, with shorter duration pulses implying physically smaller emission regions.

The location of the bright emission also can be used to distinguish emission modes.
 Giant-pulse emission is associated with gamma-ray emission and likely arises
from the outer magnetosphere \cite[][]{aaa+10a}. 
In  contrast, the bright pulses observed from Vela,
PSRs B1706$-$44, B1046$-$58,  B1133$+$16, and J1713$+$0747  are not associated with higher energy
emission but arise at the edges of the normal-mode emission.   Like PSR J0901$-$4624,  the bright pulses from PSR J0437$-$4715 arise from close to the center of the profile; however these pulses show higher polarization than weaker pulses. 

In at least some of the RRATs, emission appears to come from conventional
core-cone emission morphology across the whole polar cap \cite[][]{khv+09,kkl+11}.
In PSR~J0901$-$4624, the bright pulsars seem to originate close to
the center of the profile and hence the magnetic pole.
There is no detected  high-energy emission for PSR~J0901$-$4624, so we do not know if its bright pulses are coincident with a high-energy emitting region.

Finally, the polarization properties of these emission modes are
different. 
Both giant-pulse emission in the Crab \cite[][]{hkw+03} and giant micropulse emission in the Vela pulsar  \cite[][]{kjv02} are highly polarised.
 For the RRAT PSR~J1819$-$1458, the polarization properties are similar to that seen
in single pulse emission; the degree of linear polarization varies
and orthogonal mode jumps  are present \cite[][]{khv+09}. 
The bright pulses from PSR~J0901$-$4624 show low levels of polarization, in contrast to its normal-mode emission, which at the same longitude shows a high level of linear polarization. 

There are two plausible interpretations for the different emission modes.  The bright pulses could arise from a different plasma state or from a different region in the magnetosphere.

  Because the bright pulses originate close to the geometric center of the pulse profile, they likely originate from  low altitude, and is therefore unlikely to be depolarised by caustic effects.  Therefore the low level of polarisation suggests that the emission arises from a different plasma state, for example when the polar gap is plasma-starved.  The second state would have to produce normal-mode emission which is observed simultaneously to the bright pulses.

\section{Conclusions}

We have detected bright, narrow pulses  from PSR~J0901$-$4624. 
Similar to giant-micropulse emission observed in $4$ other young energetic pulsars, the pulse energies follow a  power-law distribution.  
Unlike other giant-micropulse emitters, the pulses originate from the center of the profile.  
Unlike the normal-mode emission for the pulsar, the bright pulses show low levels of total polarization. 
These bright pulses could possibly be produced either when the magnetosphere is in a different state or from a different location in the magnetosphere.

\section*{Acknowledgments}

We thank the referee for comments that greatly improved the manuscript.
The Parkes radio telescope is part of the Australia Telescope National 
Facility which is funded by the Commonwealth of Australia for operation as 
a National Facility managed by CSIRO. 


\end{document}